\documentclass[aps,prl,superscriptaddress,twocolumn,showpacs,floatfix]{revtex4-1}
\usepackage{graphicx}
\usepackage{dcolumn}
\usepackage{bm}

\begin{document}

\preprint{APS/123-QED}

\title{Dark Matter Search Results from the PICO-60 C$_3$F$_8$ Bubble Chamber} 

\author{C.~Amole}
\affiliation{Department of Physics, Queen's University, Kingston, K7L 3N6, Canada}

\author{M.~Ardid}
\affiliation{Departament de F\'isica Aplicada, IGIC - Universitat Polit\`ecnica de Val\`encia, Gandia 46730 Spain}

\author{I.~J.~Arnquist}
\affiliation{Pacific Northwest National Laboratory, Richland, Washington 99354, USA}

\author{D.~M.~Asner}
\affiliation{Pacific Northwest National Laboratory, Richland, Washington 99354, USA}

\author{D.~Baxter}
 \email{danielbaxter2013@u.northwestern.edu}
\affiliation{
Department of Physics and Astronomy, Northwestern University, Evanston, Illinois 60208, USA}
\affiliation{Fermi National Accelerator Laboratory, Batavia, Illinois 60510, USA}

\author{E.~Behnke}
\affiliation{Department of Physics, Indiana University South Bend, South Bend, Indiana 46634, USA}

\author{P.~Bhattacharjee}
\affiliation{Astroparticle Physics and Cosmology Division, Saha Institute
of Nuclear Physics, Kolkata, India}

\author{H.~Borsodi}
\affiliation{Department of Physics, Indiana University South Bend, South Bend, Indiana 46634, USA}

\author{M.~Bou-Cabo}
\affiliation{Departament de F\'isica Aplicada, IGIC - Universitat Polit\`ecnica de Val\`encia, Gandia 46730 Spain}

\author{P.~Campion}
\affiliation{Department of Physics, Drexel University, Philadelphia, Pennsylvania 19104, USA}

\author{G.~Cao}
\affiliation{Department of Physics, Queen's University, Kingston, K7L 3N6, Canada}

\author{C.~J.~Chen}
\affiliation{
Department of Physics and Astronomy, Northwestern University, Evanston, Illinois 60208, USA}

\author{U.~Chowdhury}
\affiliation{Department of Physics, Queen's University, Kingston, K7L 3N6, Canada}

\author{K.~Clark}
\affiliation{Department of Physics, Laurentian University, Sudbury, P3E 2C6, Canada}
\affiliation{SNOLAB, Lively, Ontario, P3Y 1N2, Canada}

\author{J.~I.~Collar}
\affiliation{Enrico Fermi Institute, KICP and Department of Physics,
University of Chicago, Chicago, Illinois 60637, USA}

\author{P.~S.~Cooper}
\affiliation{Fermi National Accelerator Laboratory, Batavia, Illinois 60510, USA}

\author{M.~Crisler}
\affiliation{Fermi National Accelerator Laboratory, Batavia, Illinois 60510, USA}
\affiliation{Pacific Northwest National Laboratory, Richland, Washington 99354, USA}

\author{G.~Crowder}
\affiliation{Department of Physics, Queen's University, Kingston, K7L 3N6, Canada}

\author{C.~E.~Dahl}
\affiliation{
Department of Physics and Astronomy, Northwestern University, Evanston, Illinois 60208, USA}
\affiliation{Fermi National Accelerator Laboratory, Batavia, Illinois 60510, USA}

\author{M.~Das}
\affiliation{Astroparticle Physics and Cosmology Division, Saha Institute
of Nuclear Physics, Kolkata, India}

\author{S.~Fallows}
\affiliation{Department of Physics, University of Alberta, Edmonton, T6G 2E1, Canada}

\author{J.~Farine}
\affiliation{Department of Physics, Laurentian University, Sudbury, P3E 2C6, Canada}

\author{I.~Felis}
\affiliation{Departament de F\'isica Aplicada, IGIC - Universitat Polit\`ecnica de Val\`encia, Gandia 46730 Spain}

\author{R.~Filgas}
\affiliation{Institute of Experimental and Applied Physics, Czech Technical University in Prague, Prague, Cz-12800, Czech Republic}

\author{F.~Girard}
\affiliation{Department of Physics, Laurentian University, Sudbury, P3E 2C6, Canada}
\affiliation{D\'epartement de Physique, Universit\'e de Montr\'eal, Montr\'eal, H3C 3J7, Canada}

\author{G.~Giroux}
 \email{ggiroux@owl.phy.queensu.ca}
\affiliation{Department of Physics, Queen's University, Kingston, K7L 3N6, Canada}

\author{J.~Hall}
\affiliation{Pacific Northwest National Laboratory, Richland, Washington 99354, USA}

\author{O.~Harris}
\affiliation{Department of Physics, Indiana University South Bend, South Bend, Indiana 46634, USA}
\affiliation{Northeastern Illinois University, Chicago, Illinois 60625, USA}

\author{E.~W.~Hoppe}
\affiliation{Pacific Northwest National Laboratory, Richland, Washington 99354, USA}

\author{M.~Jin}
\affiliation{
Department of Physics and Astronomy, Northwestern University, Evanston, Illinois 60208, USA}

\author{C.~B.~Krauss}
\affiliation{Department of Physics, University of Alberta, Edmonton, T6G 2E1, Canada}

\author{M.~Laurin}
\affiliation{D\'epartement de Physique, Universit\'e de Montr\'eal, Montr\'eal, H3C 3J7, Canada}

\author{I.~Lawson}
\affiliation{Department of Physics, Laurentian University, Sudbury, P3E 2C6, Canada}
\affiliation{SNOLAB, Lively, Ontario, P3Y 1N2, Canada}

\author{A.~Leblanc}
\affiliation{Department of Physics, Laurentian University, Sudbury, P3E 2C6, Canada}

\author{I.~Levine}
\affiliation{Department of Physics, Indiana University South Bend, South Bend, Indiana 46634, USA}

\author{W.~H.~Lippincott}
\affiliation{Fermi National Accelerator Laboratory, Batavia, Illinois 60510, USA}

\author{F.~Mamedov}
\affiliation{Institute of Experimental and Applied Physics, Czech Technical University in Prague, Prague, Cz-12800, Czech Republic}

\author{D.~Maurya}
\affiliation{Bio-Inspired Materials and Devices Laboratory (BMDL), Center for Energy Harvesting Material and Systems (CEHMS), Virginia Tech, Blacksburg, Virginia 24061, USA}

\author{P.~Mitra}
\affiliation{Department of Physics, University of Alberta, Edmonton, T6G 2E1, Canada}

\author{T.~Nania}
\affiliation{Department of Physics, Indiana University South Bend, South Bend, Indiana 46634, USA}

\author{R.~Neilson}
\affiliation{Department of Physics, Drexel University, Philadelphia, Pennsylvania 19104, USA}

\author{A.~J.~Noble}
\affiliation{Department of Physics, Queen's University, Kingston, K7L 3N6, Canada}

\author{S.~Olson}
\affiliation{Department of Physics, Queen's University, Kingston, K7L 3N6, Canada}

\author{A.~Ortega}
\affiliation{Enrico Fermi Institute, KICP and Department of Physics,
University of Chicago, Chicago, Illinois 60637, USA}

\author{A.~Plante}
\affiliation{D\'epartement de Physique, Universit\'e de Montr\'eal, Montr\'eal, H3C 3J7, Canada}

\author{R.~Podviyanuk}
\affiliation{Department of Physics, Laurentian University, Sudbury, P3E 2C6, Canada}

\author{S.~Priya}
\affiliation{Bio-Inspired Materials and Devices Laboratory (BMDL), Center for Energy Harvesting Material and Systems (CEHMS), Virginia Tech, Blacksburg, Virginia 24061, USA}

\author{A.~E.~Robinson}
\affiliation{Fermi National Accelerator Laboratory, Batavia, Illinois 60510, USA}

\author{A.~Roeder}
\affiliation{Department of Physics, Indiana University South Bend, South Bend, Indiana 46634, USA}

\author{R.~Rucinski}
\affiliation{Fermi National Accelerator Laboratory, Batavia, Illinois 60510, USA}

\author{O.~Scallon}
\affiliation{Department of Physics, Laurentian University, Sudbury, P3E 2C6, Canada}

\author{S.~Seth}
\affiliation{Astroparticle Physics and Cosmology Division, Saha Institute
of Nuclear Physics, Kolkata, India}

\author{A.~Sonnenschein}
\affiliation{Fermi National Accelerator Laboratory, Batavia, Illinois 60510, USA}

\author{N.~Starinski}
\affiliation{D\'epartement de Physique, Universit\'e de Montr\'eal, Montr\'eal, H3C 3J7, Canada}

\author{I.~\v{S}tekl}
\affiliation{Institute of Experimental and Applied Physics, Czech Technical University in Prague, Prague, Cz-12800, Czech Republic}

\author{F.~Tardif}
\affiliation{D\'epartement de Physique, Universit\'e de Montr\'eal, Montr\'eal, H3C 3J7, Canada}

\author{E.~V\'azquez-J\'auregui}
\affiliation{Instituto de F\'isica, Universidad Nacional Aut\'onoma de M\'exico, M\'exico D. F. 01000, M\'exico}
\affiliation{Department of Physics, Laurentian University, Sudbury, P3E 2C6, Canada}

\author{J.~Wells}
\affiliation{Department of Physics, Indiana University South Bend, South Bend, Indiana 46634, USA}

\author{U.~Wichoski}
\affiliation{Department of Physics, Laurentian University, Sudbury, P3E 2C6, Canada}

\author{Y.~Yan}
\affiliation{Bio-Inspired Materials and Devices Laboratory (BMDL), Center for Energy Harvesting Material and Systems (CEHMS), Virginia Tech, Blacksburg, Virginia 24061, USA}

\author{V.~Zacek}
\affiliation{D\'epartement de Physique, Universit\'e de Montr\'eal, Montr\'eal, H3C 3J7, Canada}

\author{J.~Zhang}
\affiliation{
Department of Physics and Astronomy, Northwestern University, Evanston, Illinois 60208, USA}

\collaboration{PICO Collaboration}

\date{\today}

\begin{abstract}

New results are reported from the operation of the PICO-60 dark matter detector, a bubble chamber filled with 52 kg of C$_3$F$_8$ located in the SNOLAB underground laboratory. As in previous PICO bubble chambers, PICO-60 C$_3$F$_8$ exhibits excellent electron recoil and alpha decay rejection, and the observed multiple-scattering neutron rate indicates a single-scatter neutron background of less than 1 event per month. A blind analysis of an efficiency-corrected 1167-kg-day exposure at a 3.3-keV thermodynamic threshold reveals no single-scattering nuclear recoil candidates, consistent with the predicted background. These results set the most stringent direct-detection constraint to date on the WIMP-proton spin-dependent cross section at 3.4 $\times$ 10$^{-41}$~cm$^2$ for a 30-GeV$\thinspace$c$^{-2}$ WIMP, more than one order of magnitude improvement from previous PICO results.

\end{abstract}

\pacs{29.40.-n, 95.35.+d, 95.30.Cq, FERMILAB-PUB-17-058-AE-PPD}
                          
\maketitle

The evidence for nonbaryonic dark matter in the galactic halo is compelling~\cite{PDG,dmevidence}. Many classes of theory, including supersymmetric extensions to the Standard Model (SUSY), provide promising dark matter candidates in the form of non-relativistic, weakly interacting, massive particles (WIMPs)~\cite{Jungman}. The search for WIMPs is challenging due to the predicted small WIMP-nucleon scattering cross section and nuclear recoil energies in the range of 1 to 100~keV. Low thresholds, large exposures, and background suppression are therefore critical to obtain sufficient sensitivity. As the nature of the WIMP-nucleon interaction is unknown, explorations in both the spin-dependent (SD) and spin-independent (SI) couplings are essential~\cite{wimpdetection,Snowmass,wimptheory}.

The PICO collaboration searches for WIMPs using superheated bubble chambers operated in thermodynamic conditions at which they are virtually insensitive to gamma or beta radiation. Further background suppression is achieved through the measurement of the bubble's acoustic emission, allowing for discrimination between signals from alpha decays and those from nuclear recoils~\cite{PICASSOdiscrimination}. Superheated detectors filled with fluorine-rich liquids have consistently provided the strongest constraints to spin-dependent WIMP-proton interactions~\cite{2l_13,2l_15,30l_13,PRD,previousPRL,PICASSOlimit,PICASSOFinallimit,SIMPLE}. Our largest bubble chamber to date, PICO-60, was recently filled with a 52.2 $\pm$ 0.5~kg C$_3$F$_8$ target, and operated at SNOLAB in Sudbury, Ontario, Canada. Here we report results from the first run of PICO-60 with C$_3$F$_8$, with an efficiency-corrected dark matter exposure of 1167~kg-days, taken between November 2016 and January 2017.

The PICO Collaboration previously reported the observation of anomalous background events in dark matter search data with the 2-liter PICO-2L C$_3$F$_8$~\cite{2l_13} and the 18-liter PICO-60 CF$_3$I~\cite{30l_13} bubble chambers. Improvements in fluid handling and bubble chamber operation eliminated this anomalous background in a second run of the PICO-2L detector~\cite{2l_15}. A leading hypothesis for the cause of these background events is bubble nucleation due to surface tension effects introduced by the contamination of the active target with particulate matter and water droplets~\cite{alan}. The PICO-60 detector was recommissioned following a rigorous cleaning procedure targeting particulate contamination. Every component was cleaned to MIL-STD-1246 Level 50~\cite{1246C} prior to assembly, and samples of the water buffer were taken using an \textit{in situ} filtration system during commissioning to monitor particulate injection. A final measurement after C$_{3}$F$_{8}$ distillation confirmed that the total assembly met MIL-STD-1246 Level 100, after which the inner volume was closed.

The PICO-60 apparatus was described in Ref.~\cite{30l_13}, and here we restrict ourselves to describing subsequent improvements and changes. A new seal design was deployed between the silica jar and the stainless steel bellows to minimize particulate generation, replacing the gold wire seal described in Ref.~\cite{30l_13} with a non-expanded virgin PTFE gasket. The C$_3$F$_8$ target does not require the addition of chemicals to remove free ions, unlike CF$_3$I. While the same water tank is used, a new chiller system holds the temperature in the water tank uniform to approximately 0.1$^{\circ}$C. The target volume was more than doubled, requiring a corresponding increase from two to four cameras (in two vertical columns). Eight piezoelectric acoustic transducers identical to those used in Ref.~\cite{2l_15} were attached, evenly-spaced around the outside of the silica jar, using a spring loaded HDPE ring. Five sensors failed during commissioning, leaving three operable sensors for the duration of the experiment.

The chamber expansion cycle is similar to that employed in the previous run~\cite{30l_13}. First, the chamber pressure is lowered to a predetermined point, superheating the C$_3$F$_8$ active liquid and putting our detector in a live, or expanded, state. Energy deposition within the superheated liquid will nucleate a phase change which can lead to a macroscopic gas bubble, or event. The primary trigger uses the change in entropy between two consecutive camera images~\cite{shannon} to detect the appearance of a gas bubble in the chamber. A trigger is also sent if a rise in pressure is detected or when the chamber has been expanded for 2000~s. Following a trigger, the hydraulic system initiates a fast compression, raising the pressure above 150~psia in roughly 100~ms.  The chamber begins a new expansion after a compressed deadtime of 100~s. A long compression of 600~s is imposed on every tenth compression or after a pressure-rise trigger. Of the 44.6~days of detector operation during the WIMP search dataset, the chamber was expanded (live) for 34.3~days after compressed deadtime is removed.

The WIMP search dataset was taken at 30.2 $\pm$ 0.3~psi and 13.9 $\pm$ 0.1~$^{\circ}$C. The threshold is calculated from these thermodynamic conditions using Equation 2 of Ref.~\cite{30l_13} to be 3.29 $\pm$ 0.09~keV. There is an additional 0.2~keV uncertainty in the threshold due to the thermodynamic properties of C$_{3}$F$_{8}$ taken from Ref.~\cite{REFPROP}. As discussed in Refs.~\cite{2l_13,30l_13}, the nuclear recoil threshold is not a step function at the calculated thermodynamic threshold due to energy losses that escape the region of bubble formation. \textit{In situ} nuclear and electronic recoil calibrations were performed by exposing the chamber to AmBe and $^{252}$Cf neutron sources and a $^{133}$Ba gamma source both before and after the WIMP search run. Pre-physics background data were taken during commissioning to measure the alpha backgrounds due to $^{222}$Rn chain decays which, event-by-event, are indistinguishable from nuclear recoils except in acoustic response. For the WIMP search run, we performed a blind analysis by masking the acoustic information that allows the discrimination between alpha decays and nuclear recoils, effectively salting our WIMP search data with single bulk bubbles. This information was processed only after cuts and efficiencies for single bulk nuclear recoil candidates were set, using source calibrations and pre-physics background data.

\begin{figure}
\includegraphics[width=240 pt,trim=0 0 0 0,clip=true]{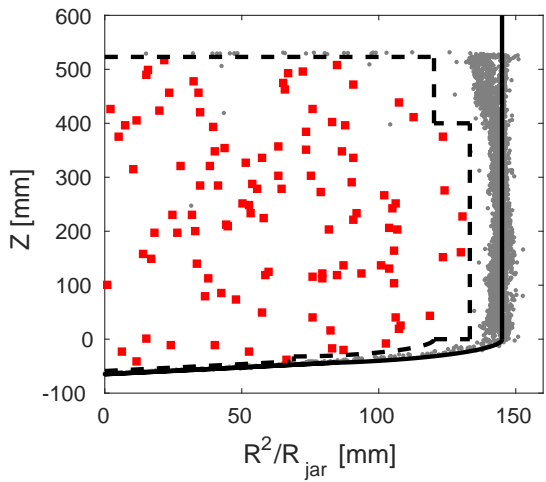}
\caption{\label{fig:XYZ} Spatial distribution of single-bubble events in the WIMP search data. Z is the reconstructed vertical position of the bubble, and R$^{2}$/R$_{\mathrm{jar}}$ is the distance from the center axis squared, normalized by the nominal jar radius (145~mm). The fiducial cut is represented by the dashed line. Red squares are the 106 single bulk bubbles passing all cuts prior to acoustic unblinding and grey dots are all rejected single-bubble events.}
\end{figure}

For the WIMP search dataset, periods of unstable operation are removed, these being defined as times within one hour of radioactive source transport near the detector or in a 24-hour window following any significant interruption to operation. The first 25~s of every expansion is discarded to remove transient effects. Of the 34.3~days the detector was expanded, 30.0~live-days (87.4$\%$) are considered in the WIMP search.

\begin{table*}
\begin{center}
\begin{tabular*}{\textwidth}{  l @{\extracolsep{\fill}} c c c c }
\hline 
\hline
\rule{0pt}{2.5ex}Dataset & Efficiency ($\%$) & Fiducial Mass (kg) & Exposure (kg-days) & No. of events \\
\hline
\rule{0pt}{2.5ex}Singles & 85.1 $\pm$ 1.8 & 45.7 $\pm$ 0.5 & 1167 $\pm$ 28 & 0 \\
\rule{0pt}{2.5ex}Multiples & 99.4 $\pm$ 0.1 & 52.2 $\pm$ 0.5 & 1555 $\pm$ 15 & 3 \\ \hline
\hline
\end{tabular*}
\caption{Summary of the final number of events and exposure determination for singles and multiples in the 30.0 live-day WIMP search dataset of PICO-60 C$_{3}$F$_{8}$ at 3.3 keV thermodynamic threshold.}
\label{table:runinfo}
\end{center}
\end{table*}

Bubble images are identified using the same entropy algorithm as used for the optical trigger. The pixel coordinates are then reconstructed into spatial coordinates using ray propagation in a simulated optical geometry. The fiducial volume is determined by setting cut values on isolated wall and surface event distributions in the source calibration and pre-physics background datasets, and is shown in Fig.~\ref{fig:XYZ}. These cuts remove events on or near the surface or within 6~mm of the nominal wall location. For regions of the detector where the optics are worse, such as the transition to the lower hemisphere, the outer 13~mm are removed. The fiducial cuts accept a mass of 45.7 $\pm$ 0.5~kg, or 87.7$\%$ of the total C$_3$F$_8$ mass.

The first step in the WIMP candidate selection removes events that are written improperly on disk, events that were not triggered by the cameras, and events for which the pressure was more than 1 psi from the target pressure. The signal acceptance for these cuts is greater than 99.9$\%$. Only events that are optically reconstructed as a single bubble are selected as WIMP candidates. This cut removes neutron-induced multiple-bubble events and events for which the optical reconstruction failed. The acceptance of this cut is 98.0 $\pm$ 0.5$\%$. In addition to the optical reconstruction fiducial cut, fiducial-bulk candidates are selected based on a rate-of-pressure-rise measurement, which is found to accept all optically reconstructed single bulk bubbles in the source calibration data.

The acoustic analysis is similar to the procedure described in~\cite{PRD} to calculate the Acoustic Parameter (AP), a measurement of the bubble's nucleation acoustic energy. As AP is used to discriminate alpha particles from nuclear recoils, events with high pre-trigger acoustic noise or an incorrectly reconstructed signal start time are removed from the WIMP candidates selection. The efficiency for these acoustic quality cuts is 99.6 $\pm$ 0.2$\%$. For this analysis, based on the pre-physics background and calibration data, AP is found to optimally discriminate alpha particles from nuclear recoils using the signals of two out of the three working acoustic transducers in the 55~kHz to 120~kHz frequency range. The AP distribution for nuclear recoil events is normalized to 1 based on AmBe and $^{252}$Cf nuclear recoil calibration data. 

An additional metric, NN score, is constructed from the piezo traces using a neural network~\cite{matlab} trained to distinguish pure alpha events (NN score = 1) from pure nuclear or electron recoil events (NN score = 0). The two-layer feedforward network takes as an input the bubble's 3D position and the noise-subtracted acoustic energy of each of three working acoustic transducers in 8 frequency bands ranging from 1~kHz to~300 kHz. The network is trained and validated with source calibration data and the pre-physics background data. A nuclear recoil candidate is defined as having an AP between 0.5 to 1.5 and a NN score less than 0.05. These combined acoustic cuts are determined to have an acceptance of 88.5 $\pm$ 1.6$\%$ based on neutron calibration single bulk bubbles. 

\begin{figure}
\includegraphics[width=240 pt,trim=0 0 0 0,clip=true]{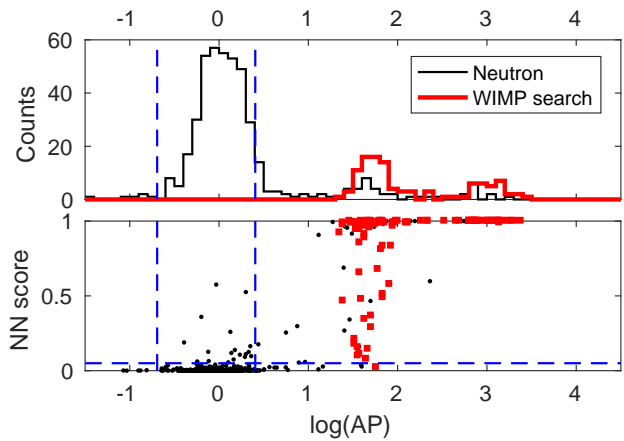}
\caption{\label{fig:APvsNN} Top: AP distributions for AmBe and $^{252}$Cf neutron calibration data (black) and WIMP search data (red) at 3.3~keV threshold. Bottom: AP and NN score for the same dataset. The acceptance region for nuclear recoil candidates, defined before WIMP search acoustic data unmasking using neutron calibration data, are displayed with dashed lines and reveal no candidate events in the WIMP search data. Alphas from the $^{222}$Rn decay chain can be identified by their time signature and populate the two peaks in the WIMP search data at high AP. Higher energy alphas from $^{214}$Po are producing larger acoustic signals.}
\end{figure}

In the WIMP search data, before unmasking acoustic information, all single bulk bubbles are identified and manually scanned. Any events with mismatched pixel coordinates are discarded. The same procedure is found to keep 98.7 $\pm$ 0.7$\%$ of single bulk bubbles in the neutron calibration data. A total of 106 single bulk bubbles pass all cuts prior to acoustic unblinding and are shown in Fig.~\ref{fig:XYZ}. The unmasking of the acoustic data, performed after completion of the WIMP search run, reveals that none of the identified 106 single bulk bubbles are consistent with the nuclear recoil hypothesis defined by AP and NN score, as shown in Fig.~\ref{fig:APvsNN}. Instead, all 106 single bulk bubbles are alpha-like in their acoustic response. The final efficiencies and exposures for the WIMP search are summarized in Table~\ref{table:runinfo}.

Neutrons produced by ($\alpha$,n) and spontaneous fission from $^{238}$U and $^{232}$Th characteristically scatter multiple times in the detector. The multiple-bubble events are an unambiguous signature and provide a measurement of the neutron background. To isolate multiple-bubble events in the WIMP search data, we do not apply acoustic or fiducial cuts, resulting in the larger exposure shown in Table~\ref{table:runinfo}. Instead, given 99.5 $\pm$ 0.1$\%$ efficiency to reconstruct at least one bubble in the bulk for a multiple-bubble event, every passing event is scanned for multiplicity. This scan reveals 3 multiple-bubble events in the WIMP search dataset. Based on a detailed Monte Carlo simulation, the background from neutrons is predicted to be 0.25 $\pm$ 0.09 (0.96 $\pm$ 0.34) single(multiple)-bubble events. PICO-60 was exposed to a 1~mCi $^{133}$Ba source both before and after the WIMP search data, which, compared against a Geant4~\cite{GEANT4} Monte Carlo simulation, gives a measured nucleation efficiency for electron recoil events above 3.3~keV of (1.80 $\pm$ 0.38)$\times$10$^{-10}$. Combining this with a Monte Carlo simulation of the external gamma flux from~\cite{gamma,alan}, we predict 0.026 $\pm$ 0.007 events due to electron recoils in the WIMP search exposure. The background from coherent scattering of $^8$B solar neutrinos is calculated to be 0.055 $\pm$ 0.007 events. 

\begin{figure}
\includegraphics[width=240 pt,trim=0 0 25 15,clip=true]{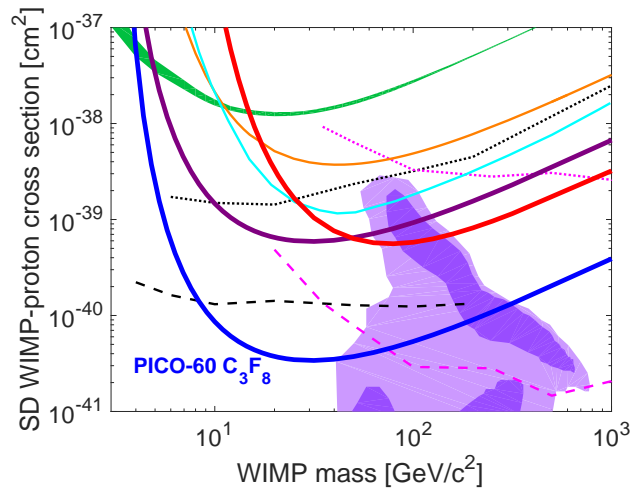}
\caption{\label{fig:SD} The  90\% C.L. limit on the SD WIMP-proton cross section from PICO-60 C$_3$F$_8$ plotted in thick blue, along with limits from PICO-60 CF$_3$I (thick red)~\cite{30l_13}, PICO-2L (thick purple)~\cite{2l_15}, PICASSO (green band)~\cite{PICASSOFinallimit}, SIMPLE (orange)~\cite{SIMPLE}, PandaX-II (cyan)~\cite{PANDAX-II}, IceCube (dashed and dotted pink)~\cite{ICECUBElimit}, and SuperK (dashed and dotted black)~\cite{SKlimit,SKlimit2}. The indirect limits from IceCube and SuperK assume annihilation to $\tau$ leptons (dashed) and {\it b} quarks (dotted). The purple region represents parameter space of the constrained minimal supersymmetric model of~\cite{SDblob}. Additional limits, not shown for clarity, are set by LUX~\cite{LUX_SD} and XENON100~\cite{XENON100} (comparable to PandaX-II) and by ANTARES~\cite{Ant1,Ant2} (comparable to IceCube).}
\end{figure}

\begin{figure}
\includegraphics[width=240 pt,trim=0 0 25 15,clip=true]{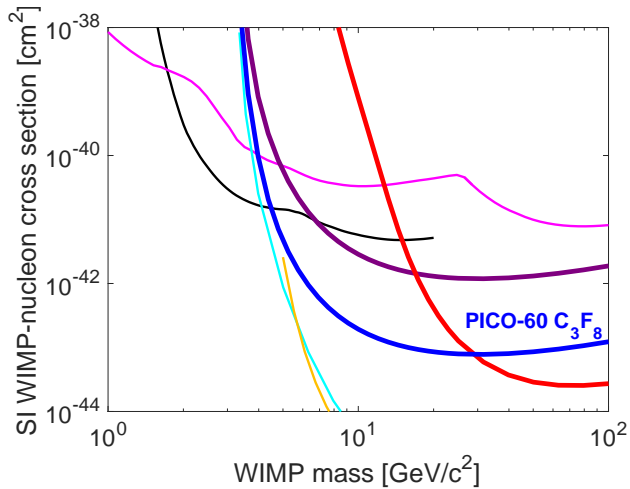}
\caption{\label{fig:SI} The 90\% C.L. limit on the SI WIMP-nucleon cross-section from PICO-60 C$_3$F$_8$ plotted in thick blue, along with limits from PICO-60 CF$_3$I (thick red)~\cite{30l_13}, PICO-2L (thick purple)~\cite{2l_15}, LUX (yellow)~\cite{LUX2017}, PandaX-II (cyan)~\cite{PANDAX-II_SI}, CRESST-II (magenta)~\cite{CRESST}, and CDMS-lite (black)~\cite{CDMSlite}. While we choose to highlight this result, LUX sets the strongest limits on WIMP masses greater than 6 GeV/c$^2$. Additional limits, not shown for clarity, are set by PICASSO~\cite{PICASSOFinallimit}, XENON100~\cite{XENON100}, DarkSide-50~\cite{DarkSide50}, SuperCDMS~\cite{SuperCDMS}, CDMS-II~\cite{CDMSII}, and Edelweiss-III~\cite{Edelweiss}.}
\end{figure}

We use the same shapes of the nucleation efficiency curves for fluorine and carbon nuclear recoils as found in Ref.~\cite{2l_13}, rescaled upwards in recoil energy to account for the 2$\%$ difference in thermodynamic threshold. We adopt the standard halo parametrization~\cite{lewinandsmith}, with the following parameters:  $\rho_D$=0.3~GeV$\thinspace$c$^{-2}$$\thinspace$cm$^{-3}$, $v_{\mathrm{esc}}$ = 544~km/s, $v_{\mathrm{Earth}}$ = 232~km/s, and $v_o$ = 220~km/s. We use the effective field theory treatment and nuclear form factors described in Refs.~\cite{spindependentcouplings,Anand,Gresham,Gluscevic} to determine sensitivity to both spin-dependent and spin-independent dark matter interactions. For the SI case, we use the $M$ response of Table 1 in Ref.~\cite{spindependentcouplings}, and for SD interactions, we use the sum of the $\Sigma^{\prime}$ and $\Sigma^{\prime \prime}$ terms from the same table. To implement these interactions and form factors, we use the publicly available \texttt{dmdd} code package~\cite{Gluscevic,Gluscevic2}. The calculated Poisson upper limits at the 90$\%$~C.L. for the spin-dependent WIMP-proton and spin-independent WIMP-nucleon elastic scattering cross-sections, as a function of WIMP mass, are shown in Fig.~\ref{fig:SD} and~\ref{fig:SI}. These limits, corresponding to an upper limit on the spin-dependent WIMP-proton cross section of 3.4 $\times$ 10$^{-41}$~cm$^2$ for a 30~GeV$\thinspace$c$^{-2}$ WIMP, are currently the world-leading constraints in the WIMP-proton spin-dependent sector and indicate an improved sensitivity to the dark matter signal of a factor of 17, compared to previously reported PICO results.

\begin{figure}
\includegraphics[width=240 pt,trim=0 0 0 0,clip=true]{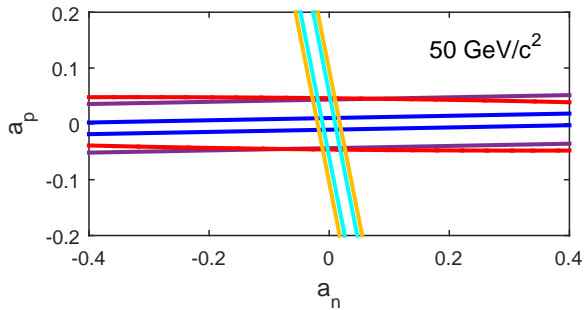}
\caption{\label{fig:anap} PICO-60 constraints (blue) on the effective spin-dependent WIMP-proton and WIMP-neutron couplings, $a_p$ and $a_n$, for a 50~GeV/c$^2$ WIMP mass. Parameter space outside of the band is excluded. Also shown are results from PANDAX-II (cyan)~\cite{PANDAX-II}, LUX (yellow)~\cite{LUX_SD}, PICO-2L (purple)~\cite{2l_15}, and PICO-60 C$_3$FI (red)~\cite{30l_13}.}
\end{figure}

A comparison of our proton-only SD limits with neutron-only SD limits set by other dark matter search experiments is achieved by setting constraints on the effective spin-dependent WIMP-neutron and WIMP-proton couplings $a_n$ and $a_p$ that are calculated according to the method proposed in Ref.~\cite{Tovey}. The expectation values for the proton and neutron spins for the $^{19}$F nucleus are taken from Ref.~\cite{spindependentcouplings}. The allowed region in the $a_n-a_p$ plane is shown for a 50~GeV$\thinspace$c$^{-2}$ WIMP in Fig.~\ref{fig:anap}. We find that PICO-60 C$_3$F$_8$ improves the constraints on $a_n$ and $a_p$, in complementarity with other dark matter search experiments that are more sensitive to the WIMP-neutron coupling.

\begin{figure}
\includegraphics[width=240 pt,trim=0 0 0 0,clip=true]{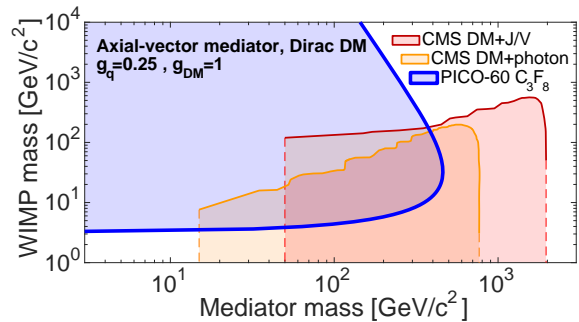}
\caption{\label{fig:Mediator} Exclusion limits at 95$\%$ C.L. in the $m_\mathrm{DM} - m_\mathrm{med}$ plane. PICO-60 constraints (thick blue) are compared against collider constraints from CMS  for an axial-vector mediator using the monojet/mono-V (red)~\cite{CMS_Exo_JV} and mono-photon (orange)~\cite{CMS_Exo_photon} channels. The shaded regions signify excluded parameter space for the chosen model. A similar analysis by ATLAS can be found in~\cite{ATLAS}.}
\end{figure}

The LHC has significant sensitivity to dark matter, but to interpret LHC searches, one must assume a specific model to generate the signal that is then looked for in the data. Despite this subtlety, the convention has been to show LHC limits alongside more general direct detection constraints in the parameter space of Fig.~\ref{fig:SD}. We choose instead to compare our limits with those of the LHC on the chosen model, as shown in Fig.~\ref{fig:Mediator}. The LHC Dark Matter Working Group has made recommendations on a set of simplified models to be used in LHC searches and the best way to present such results~\cite{Buchmueller2014, DMLHC2015, Boveia2016}. For a given simplified model involving a mediator exchanged via the $s$-channel, there are four free parameters: the dark matter mass $m_{\mathrm{DM}}$, the mediator mass $m_{\mathrm{med}}$, the universal mediator coupling to quarks $g_q$, and the mediator coupling to dark matter $g_{\mathrm{DM}}$. We make a direct comparison of the sensitivity of PICO to that of CMS~\cite{CMS_Exo_JV,CMS_Exo_photon} by applying our results to the specific case of a simplified dark matter model involving an axial-vector $s$-channel mediator. Following Eq. 4.7-4.10 of Ref.~\cite{Boveia2016}, we find an expression for the spin-dependent cross section as a function of those free parameters, and we invert this expression to find $m_\mathrm{med}$ as a function of cross section. For this comparison, we assume $g_q = 0.25$ and $g_\mathrm{DM} = 1$. With this simple translation onto the specified model, we can plot our limits on the same $m_\mathrm{DM} - m_\mathrm{med}$ plane, and the results are shown in Fig.~\ref{fig:Mediator}.

The PICO collaboration wishes to thank SNOLAB and its staff for support through underground space, logistical and technical services. SNOLAB operations are supported by the Canada Foundation for Innovation and the Province of Ontario Ministry of Research and Innovation, with underground access provided by Vale at the Creighton mine site. We are grateful to Kristian Hahn and Stanislava Sevova of Northwestern University and Bj\"orn Penning of the University of Bristol for their assistance and useful discussion. We wish to acknowledge the support of the Natural Sciences and Engineering Research Council of Canada (NSERC) and the Canada Foundation for Innovation (CFI) for funding. We acknowledge the support from National Science Foundation (NSF) (Grant 0919526, 1506337, 1242637 and 1205987). We acknowledge that this work is supported by the U.S. Department of Energy (DOE) Office of Science, Office of High Energy Physics (under award DE-SC-0012161), by the DOE Office of Science Graduate Student Research (SCGSR) award,  by DGAPA-UNAM through grant PAPIIT No. IA100316 and CONACyT (Mexico) through grant No. 252167, by the Department of Atomic Energy (DAE), the Government of India, under the Center of AstroParticle Physics II project (CAPP-II) at SAHA Institute of nuclear Physics (SINP), the Czech Ministry of Education, Youth and Sports (Grant LM2015072) and the the Spanish Ministerio de Econom\'ia y Competitividad, Consolider MultiDark (Grant CSD2009-00064). This work is partially supported by the Kavli Institute for Cosmological Physics at the University of Chicago through NSF grant 1125897, and an endowment from the Kavli Foundation and its founder Fred Kavli. We also wish to acknowledge the support from Fermi National Accelerator Laboratory under Contract No. De-AC02-07CH11359, and Pacific Northwest National Laboratory, which is operated by Battelle for the U.S. Department of Energy under Contract No. DE-AC05-76RL01830. We also thank Compute Canada (www.computecanada.ca) and the Center for Advanced Computing, ACENET, Calcul Qu\'ebec, Compute Ontario and WestGrid for the computational support.

\end{document}